\documentstyle[epsfig]{aipproc}

\title{The 4.5 $\bf \pm0.5$ Soft Gamma Repeaters in Review }
\author{K. Hurley }
\address{UC Berkeley \\ Space Sciences Laboratory \\ Berkeley, CA 94720-7450 }

\begin{document}

\maketitle

\begin{abstract}
Four Soft Gamma Repeaters (SGRs) have now been identified with certainty,
and a fifth has possibly been detected.  I will review their X-ray and gamma-ray
properties in both outburst and quiescence.  The magnetar model accounts fairly
well for the observations of SGR1806-20 and SGR1900+14, but data are still lacking
for SGR1627-41 and SGR0525-66.  The locations of the SGRs with respect to their
supernova remnants suggest that they are high velocity objects.
\end{abstract}

\section{Introduction}

The Soft Gamma Repeaters are sources of short, soft-spectrum ($\le$ 100 keV)
bursts with super-Eddington luminosities.  They undergo sporadic, unpredictable
periods of activity, sometimes quite intense, which last for days to months, often
followed by long periods (up to years or decades) during which no bursts are emitted.
Very rarely, perhaps every 20 years, they emit long duration \it giant flares \rm
which are thousands of times more energetic than the bursts, with hard spectra ($\sim$ MeV).
The SGRs are quiescent, and in some cases periodic, 1-10 keV soft X-ray sources as well.
They all appear to be associated with supernova remnants, and a good working hypothesis
is that they are all \it magnetars \rm, i.e. highly magnetized neutron stars for which
the magnetic field energy dominates all other sources, including rotation 
\cite{duncan92,thom95}.  Figure 1 shows the time histories of bursts from SGR1900+14, and figure 2 shows a typical energy spectrum.

In this paper, I will mainly review the radio, X-ray, and gamma-ray properties of the SGRs
in outburst and in quiescence, and indicate how the magnetar model accounts for these properties.

\begin{figure}[h!]
\centerline{\psfig{file=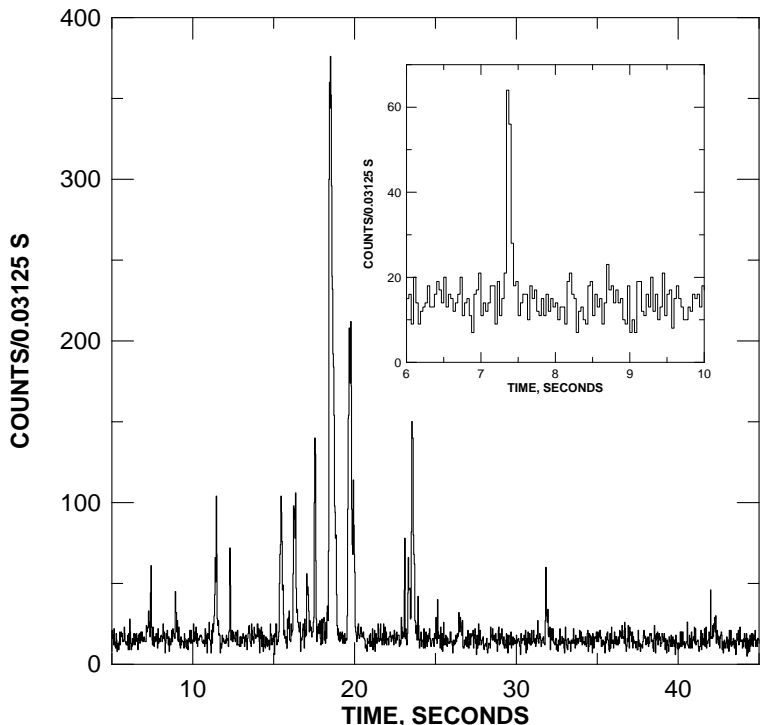, width=10cm}}
\caption{From Hurley et al. (1999a).  Inset: a typical burst from SGR1900+14
as observed in the 25-150 keV range by \it Ulysses \rm.
Main figure: bursts during a period of intense activity.}
\label{fig1}
\end{figure}

\begin{figure}
\centerline{\psfig{file=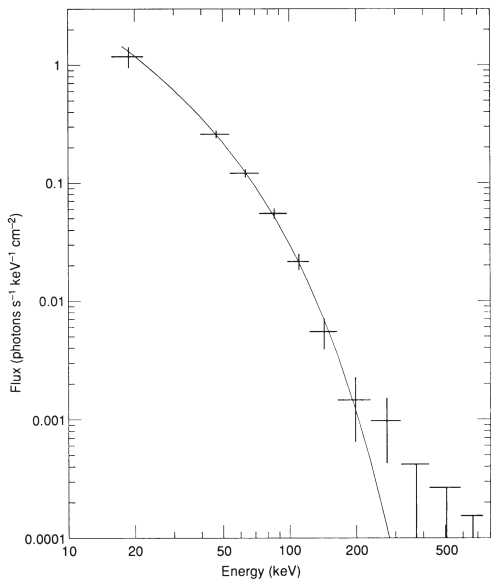, width=7cm}}
\caption{Reprinted by permission from Nature (Kouveliotou et al. 1993) copyright
1993 Macmillan Magazines Ltd..  Typical spectrum of a burst from
SGR1900+14 as observed by BATSE.  The spectrum is fit here with an optically
thin thermal bremsstrahlung function, with kT=39 keV.}
\label{fig2}
\end{figure}

\section{SGR1806-20}

Kulkarni and Frail \cite{kul93} suggested that this SGR was associated with the Galactic
supernova remnant (SNR) G10.0-0.3, based on its localization to a $\sim \rm 400 \,
 arcmin.^2$ error box by the old interplanetary network (IPN) \cite{atteia87}.
This was confirmed when ASCA observed and imaged the source \it in outburst \rm,
localizing it to a 1' error circle \cite{mur94}.  A quiescent soft X-ray
source was also detected by Cooke \cite{cooke93} using the ROSAT HRI.  Based on more recent
observations, Kouveliotou et al. \cite{kou98} have found that the quiescent source is
periodic (P=7.48 s) and is spinning down rapidly (\.{P}=$\rm 2.8 \times 10^{-11} s/s$).    
If this spindown is interpreted as being due entirely to magnetic dipole radiation,
the implied field strength is B=8 $\rm \times 10^{14} G$.  The 2-10 keV X-ray luminosity
of the source is $\rm 2 \times 10^{35} erg/s$, and the low energy X-ray spectrum
may be fit by a power law with index 2.2.

\begin{figure}[h!]
\centerline{\psfig{file=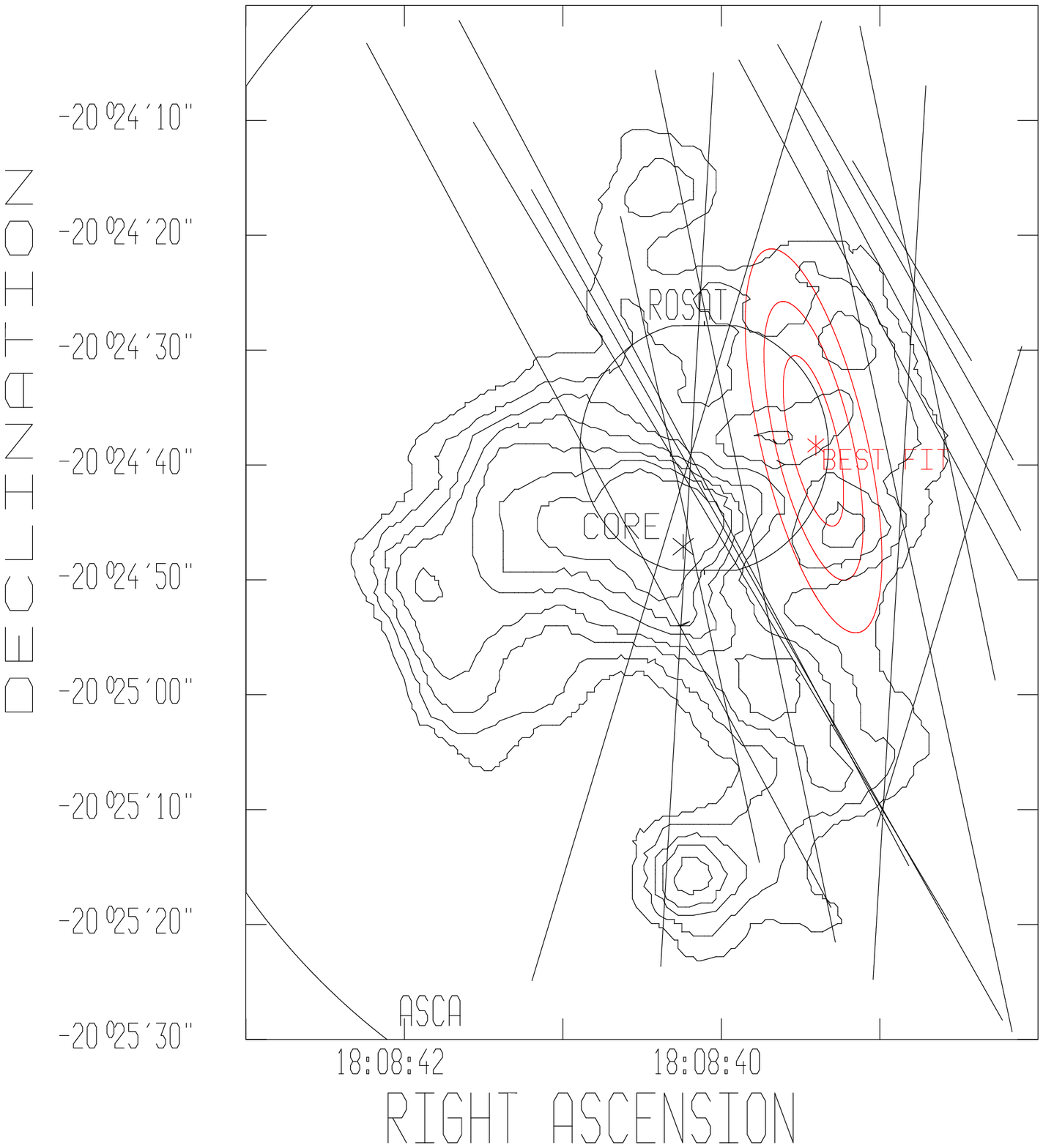, width=10cm}}
\caption{From Hurley et al. (1999b).  Eight IPN annuli (lines), and the 1, 2, and 3 $\sigma$ equivalent
confidence contours (ellipses) for SGR1806-20.  The best fit position and the position of
the non-thermal core are indicated.  The ASCA error circle is just visible in
the lower left and upper left hand corners (Murakami et al. 1994).  The ROSAT PSPC error circle is at the
center; its radius is 11" (Cooke 1993).
The 3.6 cm radio contours of G10.0-0.3 are also shown, from Vasisht et al. (1995).}
\label{fig3}
\end{figure}

The SNR G10.0-0.3 has a non-thermal core, and Frail et al. \cite{frail97} have detected
changes in the radio contours of the core on $\sim$ year timescales.  Van Kerkwijk et al.
\cite{vanK95} have found an unusual star at the center of this core, which they identify
as a luminous blue variable (LBV).  The presence of this object has been a mystery
up to now, because it was thought that the SGRs were single neutron stars.  Recent
work from the 3rd IPN has shed some light on this issue \cite{hu99b}.  
Figure 3 shows the location of the SGR superimposed on the radio
contours of the SNR.  It can be seen that the SGR is in fact offset from the LBV.
The LBV may be powering the non-thermal core of the SNR, and causing the changes
in the radio contours.  It is also possible that the SGR progenitor was once
bound to the LBV, but that it became unbound when it exploded as a supernova.
A transverse velocity of $\sim$100 km/s would then be required to explain the
displacement between the two.  Alternatively, it is possible that the apparent
SGR-SNR association is due to a chance alignment of these two objects along
the line of sight.

\section{SGR1900+14}
\begin{figure}[h!]
\centerline{\psfig{file=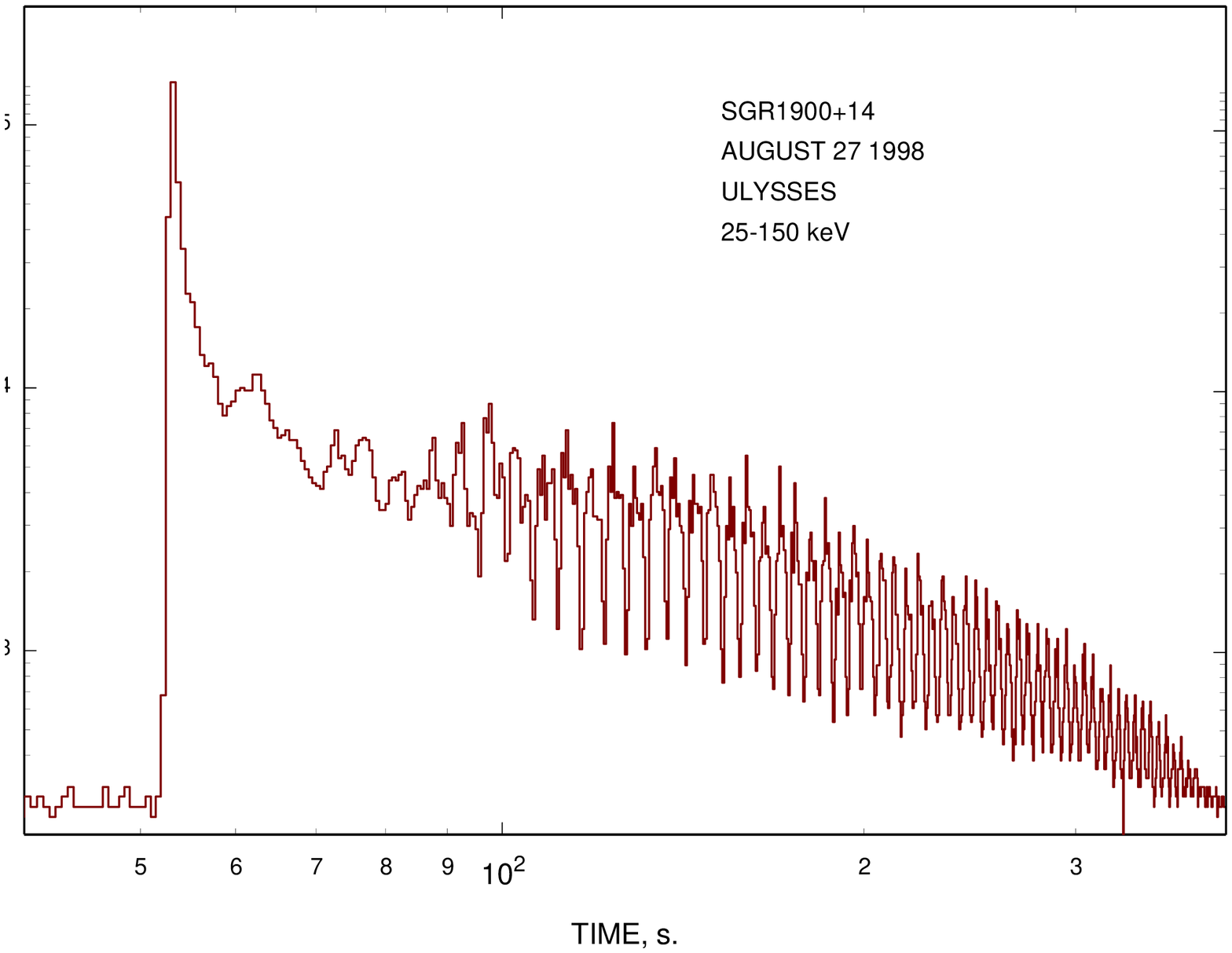, width=9cm}}
\caption[]{Reprinted by permission from Nature (Hurley et al. 1999d) copyright
1999 Macmillan Magazines Ltd..  The \it Ulysses \rm 25-150 keV time history of the
1998 August 27 giant flare from SGR1900+14.  Note the 5.16 s periodicity.}
\label{fig4}
\end{figure}

SGR1900+14 was discovered by Mazets et al. \cite{maz79} when it burst 3 times in two days.
A precise localization by the IPN \cite{hu99a} showed that this source lay just
outside the Galactic SNR G42.8+0.6, with an implied proper motion $>$1000 km/s.  The SGR
is associated with a quiescent soft X-ray source \cite{vas94,hu99c,kou99}.  
The quiescent source has a period 5.16 s, and a period derivative
6.1 $\rm \times 10^{-11} s/s$; again, assuming purely dipole radiation, B $\rm \sim
8 \times 10^{14} G$.  The 2-10 keV luminosity is 3 $\rm \times 10^{34} erg/s$, and the
spectrum may be fit with a power law of index 2.2.

On 1998 August 27, the SGR emitted a giant flare which was probably the most intense burst
ever detected at Earth \cite{hu99d}.  Its luminosity was 2 $\rm \times 10^{43}$
erg/s in $>$25 keV X-rays, or 10$^5 L_E$ (the Eddington luminosity).  The time history of this
burst clearly displayed the 5.16 s periodicity of the quiescent source (figure 4).
The magnetic field strength required to contain the electrons responsible for the
X-ray emission is $\rm >10^{14} G$; this constitutes an independent argument for the
presence of strong fields in SGRs.  From measurements of the ionospheric disturbance
which this burst caused, Inan et al. \cite{inan99} have estimated that there must have been
one order of magnitude more energy in 3-10 keV X-rays than in $>$25 keV X-rays, bringing
the total energy to $\rm \sim 4 \times 10^{44} erg$.  Frail et al. \cite{frail99} detected
a transient radio source with the VLA at the SGR position following the giant flare.
This is the only case where a radio point source is present at an SGR position.

\section{SGR0525-66}

This SGR was discovered when it emitted the giant flare of 1979 March 5 
\cite{cline80,gol79}.  It was localized by the IPN to a 0.1 $\rm arcmin^2$ error box
within the N49 supernova remnant \cite{evans80}.  For an LMC distance of 55 kpc, this
burst had a luminosity of $\rm 5 \times 10^{44}$ erg/s in X-rays $>$50 keV, or 
$\rm 2 \times 10^6 L_E$; the total energy emitted was $\rm \sim 7 \times 10^{44} erg$ in
$>$50 keV X-rays.  The time history displayed a clear 8 s periodicity \cite{barat79}.
Both Duncan and Thompson \cite{duncan92} and Paczynski \cite{pac92} suggested a strongly magnetized neutron star as
the origin of this burst.  Although the source remained active through 1983 \cite{gol87}, 
it has not been observed to burst since then.

Rothschild et al. \cite{rot94} found a quiescent soft X-ray point source in the SGR error box with
a ROSAT HRI observation.  As no energy spectra are obtained from the HRI, the soft
X-ray luminosity can only be estimated by assuming various spectral shapes.  The
0.1-2.4 keV luminosity is in the range $\rm 10^{36} - 10^{37}$ erg/s, depending on
the assumed spectrum.  No periodicity was detected in this observation, but the
upper limit to the pulsed fraction is only 66\%.  If the age of the N49 SNR is taken to
be 5 kyr \cite{vanc92}, the implied transverse velocity of the SGR is 
several thousand km/s.  \it Chandra \rm observations of the SNR are scheduled, and
are bound to reveal more about this interesting object.

\section{SGR1627-41}

SGR1627-41 burst about 100 times in June-July 1998, and has not been observed
to burst since then.  During that period, observations by BATSE \cite{woods99a}, 
\it Ulysses \rm \cite{hu99e},
KONUS-\it Wind \rm \cite{maz99}, and RXTE \cite{smit99} led to a precise source localization.  The SGR lies near the SNR G337.0-0.1, at a distance of
$\sim$ 11 kpc.  The implied transverse velocity of the SGR is in the
range 200 - 2000 km/s.  Although no giant flare
has been observed from this source, there is a KONUS-\it Wind \rm observation
of an extremely energetic event \cite{maz99}.  The luminosity and
total energy of the burst in the $>$15 keV range were $\rm \sim 8 \times 10^{43}
erg/s \, and \sim 3 \times 10^{42} erg/s$, respectively.

\begin{figure}[t!]
\centerline{\psfig{file=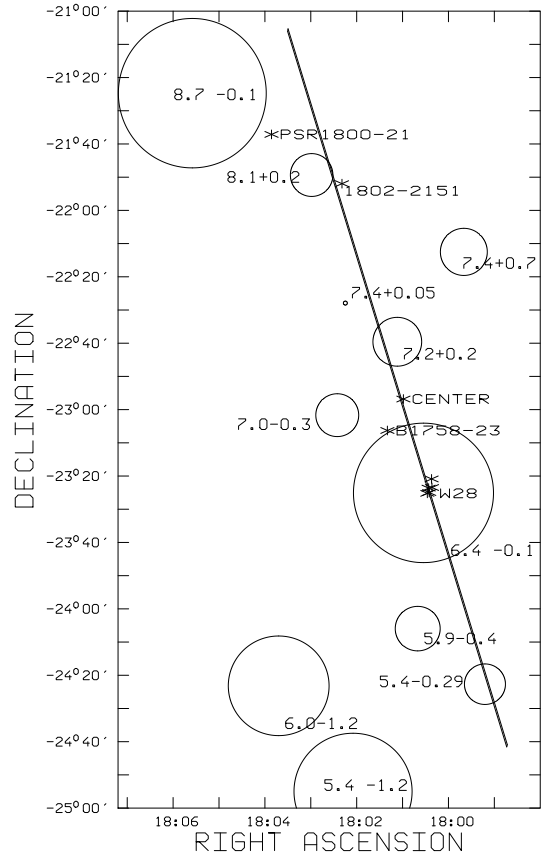, width=7cm}}
\caption[]{From Cline et al. (1999).  IPN error box for SGR1801-23 (the lines are too
closely spaced to distinguish).  The center is indicated with an asterisk.
Circles give the approximate locations of confirmed and suspected SNRs; the
radii have been taken as half the size given in the catalogs.  Asterisks give
the positions of ROSAT X-ray sources, and two pulsars, PSR1800-21 and B1758-23,
probably associated with SNRs 8.7-0.1 and 6.4-0.1. Coordinates are J2000.}
\label{fig5}
\end{figure} 

Like the other SGRs, this one also appears to be a quiescent soft X-ray source.
\it BeppoSAX \rm observations revealed a variable source with spectral index 2.1
and luminosity $\rm \sim 10^{35} erg/s$ \cite{woods99b}.  Although the
\it BeppoSAX \rm observations gave weak evidence for a possible 6.4 s periodicity,
this was not confirmed in later ASCA observations of the source with
better statistics \cite{hu99f}.

\section{SGR1801-23}

The latest SGR to be discovered is 1801-23 \cite{cline99}.  It was observed
to burst just twice, on June 29, 1997, by \it Ulysses \rm, BATSE, and KONUS-\it Wind\rm.
The burst spectra were soft, and could be fit by an optically thin thermal bremsstrahlung
function with a kT of $\sim$ 25 keV.  The time histories were short.  In both respects,
then, the source properties resemble those of the other SGRs.  However, because only
two bursts were observed, and they occurred on the same day, the IPN localization
is not very precise.  The error box is 3.8 $\rm^o$ long, and has an area of
$\rm \sim 80 \, arcmin^2$.  The source lies in the general direction of the Galactic
center, and the error box crosses numerous possible counterparts (figure 5).  The
source would have a super-Eddington luminosity for any distance $>$ 250 pc; at
the approximate distance of the Galactic center, its luminosity would be 1200L$_E$.
At present, the best hypothesis is that this source is indeed an SGR; recall that
SGR1900+14 was similarly detected when it burst 3 times in two days, and it remained
quiescent for many years.  Like SGR1900+14, the identification of SGR1801-23 may
have to await a new period of bursting activity.

Table 1 summarizes the essential properties of the SGRs.

\begin{table}[h]
\caption{Essential properties of the SGRs}
\label{table1}
\begin{tabular}{|c|cccccc}
SGR       & Super-      & Giant  & Periodicity & Quiescent  & Periodicity  & \.{P}    \\
           & Eddington   & Flare? & Observed in & Soft X-ray & in Quiescent & 10$^{-11} $ \\
           & Bursts?     &        & Burst?      & Source?    & Source?      & s/s          \\
           &             &        &             & (erg/s)    &              &           \\
\hline
1806-20 & 1000$\times$ & No     & No          & $\rm 2\times10^{35}$ & 7.47 s &  2.8  \\
1900+14 & 1000$\times$ & 270898 & 5.16 s & $\rm 3\times 10^{34}$ & 5.16 s & 6.1 \\
0525-66 & 20000$\times$ & 050379 & 8 s    & $\rm 10^{36-37}$       & No     & --- \\
1627-41 & 400000$\times$ & No        & No     & $\rm 10^{35}$          & 6.4 s? & --- \\
1801-23 &   ?             & No        & No     &   ?                         & ---    & --- \\
\end{tabular}
\end{table}

\section{The Magnetar Model}

Briefly, the magnetar model \cite{duncan92,thom95}
explains the short, soft bursts by localized cracking
on the neutron star surface, with excitation of Alfven waves which accelerate
electrons.  Every 20--100 y, a massive, global crustquake takes place.  Regions of
the neutron star with magnetic fields of opposite polarity suddenly encounter
one another, resulting in magnetic field annihilation and energization of the
magnetosphere, giving rise to a giant flare.  Magnetars are thought to be born
in $\sim$ 1 out of 10 supernova explosions, and remain active for perhaps 10,000 y.
Thus there should be about 10 active magnetars in the Galaxy at any given time.
So far, we have found 4.5 $\pm$ 0.5.  Stay tuned for more!

We are grateful for JPL support of \it Ulysses \rm operations under Contract 958056
and to NASA for support of the IPN under NAG5-7810.

\end{document}